\documentclass{article}

\usepackage{arxiv}

\usepackage[utf8]{inputenc} % allow utf-8 input
\usepackage[T1]{fontenc}    % use 8-bit T1 fonts
\usepackage{hyperref}       % hyperlinks
\usepackage{url}            % simple URL typesetting
\usepackage{booktabs}       % professional-quality tables
\usepackage{amsfonts}       % blackboard math symbols
\usepackage{nicefrac}       % compact symbols for 1/2, etc.
\usepackage{microtype}      % microtypography
\usepackage{lipsum}
\usepackage{natbib} % EV added this
\usepackage{graphicx} %EV added this too for the figure?

\title{The ASCCR Frame for Learning Essential Collaboration Skills}

\author{
  Eric A.~Vance\thanks{The authors thank our students at the University of Colorado Boulder, Virginia Tech, and Cal Poly and our mentors---especially Doug Zahn---who helped us develop and refine this material.} \\
  Department of Applied Mathematics\\
  Laboratory for Interdisciplinary Statistical Analysis (LISA)\\
  University of Colorado Boulder\\
  Boulder, CO 80309-0526 \\
  \texttt{Eric.Vance@Colorado.EDU} \\
   \And
 Heather S.~Smith \\
 Department of Statistics\\
 Cal Poly San Luis Obispo\\
 San Luis Obispo, CA 93407\\
  \texttt{hsmith@calpoly.edu} \\
}

\begin{document}
\maketitle

\begin{abstract}
Statistics and data science are especially collaborative disciplines that typically require practitioners to interact with many different people or groups. Consequently, interdisciplinary collaboration skills are part of the personal and professional skills essential for success as an applied statistician or data scientist. These skills are learnable and teachable, and learning and improving collaboration skills provides a way to enhance one's practice of statistics and data science. To help individuals learn these skills and organizations to teach them, we have developed a framework covering five essential components of statistical collaboration: Attitude, Structure, Content, Communication, and Relationship. We call this the ASCCR Frame. This framework can be incorporated into formal training programs in the classroom or on the job and can also be used by individuals through self-study. We show how this framework can be applied specifically to statisticians and data scientists to improve their collaboration skills and their interdisciplinary impact. We believe that the ASCCR Frame can help organize and stimulate research and teaching in interdisciplinary collaboration and call on individuals and organizations to begin generating evidence regarding its effectiveness.
\end{abstract}

% keywords can be removed

\section{INTRODUCTION}
\label{sec:intro}

Interdisciplinary collaboration brings two or more experts from different domains together with a common purpose and goals to achieve what neither could on their own. In particular, statistics is an inherently collaborative discipline \citep{ben-zvi_using_2007}. Applied statisticians and data scientists rarely ``own" the data they analyze, nor do they typically originate the problems to be solved and decisions to be made with data. Therefore, to have real-world impact, most applied statisticians and data scientists collaborate with various domain experts to understand and refine the questions to be answered, access and analyze the appropriate data, and present results and recommendations for decision making and implementation. One way for statisticians and data scientists to maximize the impact of their work is to improve the quality of their collaborations by learning and implementing essential collaboration skills.

The American Statistical Association (ASA); the National Academies of Science, Engineering, and Medicine (NASEM); and others have concluded that learning collaboration skills are vitally important for statistical practice for student and professional statisticians and data scientists. According to the ASA, graduates of statistical science undergraduate programs, ``Should demonstrate ability to collaborate in teams and \ldots be able to communicate complex statistical methods'' \citep[p.10]{asa_curriculum_2014}. A consensus study report of NASEM concludes, ``The ability to work well in multidisciplinary teams is a key component of data science education that is highly valued by industry. Multidisciplinary collaboration provides students with the opportunity to use creative problem solving and to refine leadership skills, both of which are essential for future project organization and management experiences in the workplace'' \citep[pp 28-29]{national_academies_of_sciences_engineering_and_medicine_data_2018}. Another NASEM consensus study report concludes, ``In an ideal STEM graduate education system \ldots Students would be encouraged to create their own project-based learning opportunities---ideally as a member of a team---as a means of developing transferable professional skills such as communication, collaboration, management, and entrepreneurship'' \citep[p. 4]{national_academies_of_sciences_engineering_and_medicine_graduate_2018}.

While learning collaboration skills is clearly essential for students in statistics and data science, it is also essential for professional statisticians. ASA states, ``Professional development is important to statisticians because it helps them advance their careers, remaining competitive and marketable \ldots Factors such as communication, leadership, and influence skills \ldots are vital to the impact of individual contributions and the visibility of our profession'' \citeyearpar[p. 1]{asa_asa_2012}. As Kettenring et al. discussed in the ``Challenges and Opportunities for Statistics in the Next 25 Years'' panel session at the 2014 Joint Statistical Meetings, ``Statistics needs to figure out how best to collaborate with computer science and other disciplines. Teamwork will be required to handle the large and difficult problems that lie ahead'' \citeyearpar[p. 89]{kettenring_challenges_2015}. B{\"o}rner et al. conclude from an analysis of trends in the increasingly data-driven economy that the demand for professional skills, like teamwork and communication, increases with greater demand for technical skills and tools \citeyearpar{borner_skill_2018}. Yet, these skills are not usually taught in the traditional statistics or computer science classroom but instead, are gained through experience and collaboration with others \citep{blei_science_2017}.

We have developed a model for learning and teaching collaboration skills. We call it the ASCCR Frame (pronounced ``ask'er'') as it covers five essential components of our theory of interdisciplinary collaboration: {\underline A}ttitude, {\underline S}tructure, {\underline C}ontent, {\underline C}ommunication, and {\underline R}elationship (see Figure 1). While this framework is general and could be applied by those who collaborate in any field, we focus specifically on applying it to collaborative statisticians and data scientists.

\begin{figure}
\begin{center}
\includegraphics[width=3in]{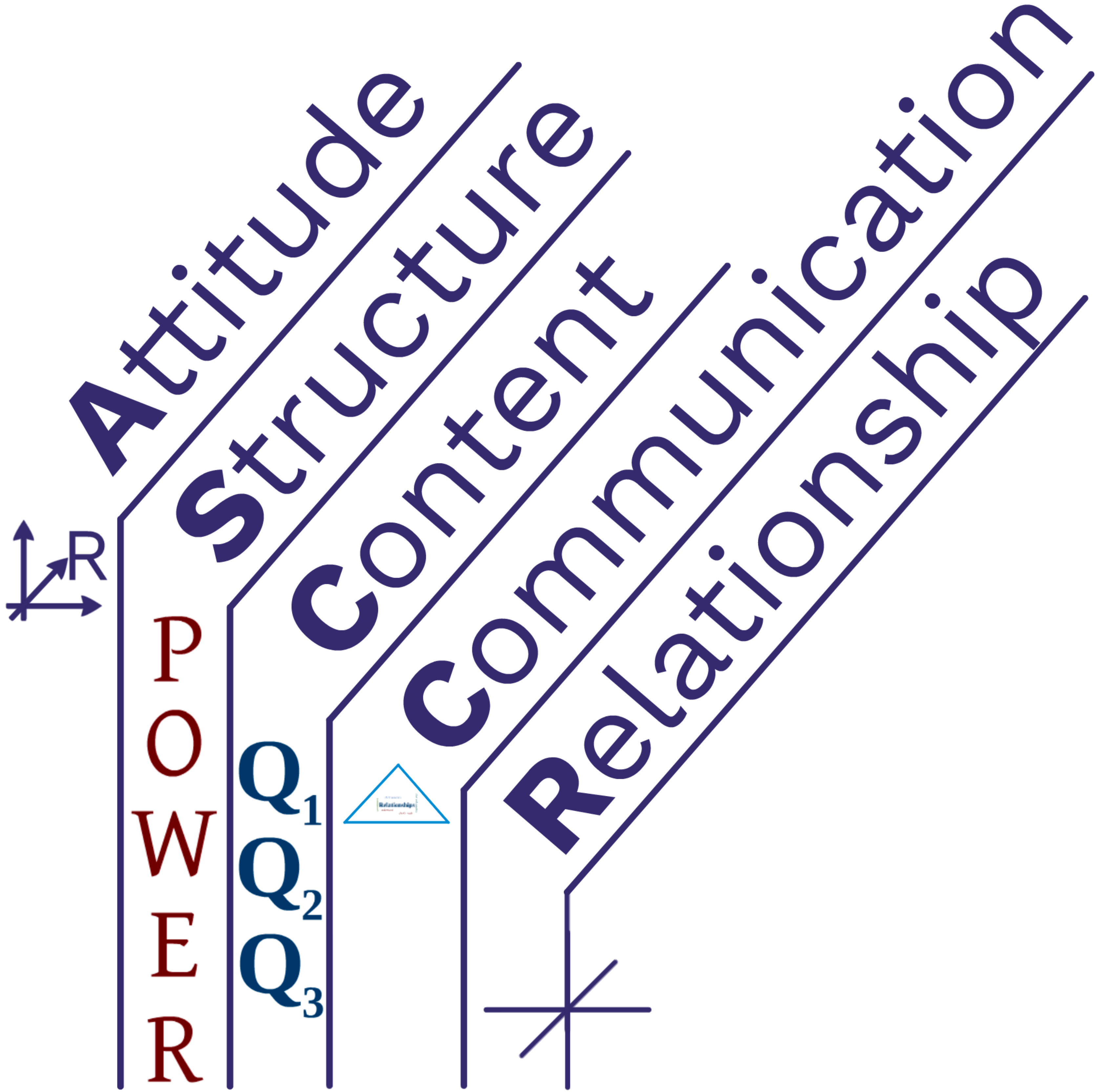}
\end{center}
\caption{The ASCCR Frame for Collaboration consisting of Attitude, Structure, Content, Communication, and Relationship. \label{fig:first}}
\end{figure}

This practical framework for interdisciplinary collaboration is based on our experience in academia as the Directors of the Laboratory for Interdisciplinary Statistical Analysis (LISA) and the Cal Poly Statistical Consulting Service. We have collectively collaborated on more than 900 projects and have supervised more than 4000 additional collaborative projects. We see firsthand the improved outcomes generated when our labs' collaborators and consultants enhance their collaboration skills. In the 2017-18 academic year, LISA had---for its first time in 10 years---a 100\% satisfaction rate from domain experts who responded to a request for feedback  \citep{vance_lisa:_2018}.

The goal of this article is to introduce the ASCCR Frame to the statistics and data science communities to help organize and stimulate research and teaching in interdisciplinary collaboration. In Sections 2-6, we highlight aspects of each individual component of the ASCCR Frame. Future work by the authors and others will provide more in-depth exploration of each component and evidence-based findings. In Section 7, we discuss how to teach ASCCR in the classroom and how both beginning and experienced statisticians can implement the ASCCR Frame to improve their collaboration skills and thus increase their impact within their organizations. We conclude this paper in Section 8 with a call for individuals and organizations to begin generating evidence regarding the effectiveness of the ASCCR Frame.

\section{{\underline A}TTITUDE}
\label{sec:attitude}
Collaboration skills are part of the personal and professional skills---along with communication, career planning, and leadership---that the American Statistical Association has deemed essential for success as a statistician or data scientist \citep{palmer_asa_2014}. From personal experience improving our own collaboration skills and extensive experience teaching students in the classroom and professionals via workshops, we observe day after day that collaboration skills are learnable and teachable. Statisticians can improve their collaboration skills to improve their effectiveness and their impact within their organizations. The mindset that these skills are important to learn and in fact can be learned is an attitude that sets the foundation for improving the collaboration skills encompassed in the ASCCR frame.

The Attitude of collaboration starts with two concepts of emotional intelligence: self-awareness and self-management \citep{goleman_emotional_2012}. Collaborative statisticians should be aware of their attitudes and emotions and should choose to act on attitudes that enhance their effectiveness as collaborators and be mindful of---but refrain from acting on---attitudes and thoughts that detract from collaboration.

Attitudes we have found enhance collaboration include:
\begin{itemize}
\item {\it The Fundamental Law of Collaboration}, adopted from \citet{covey_seven_1989}: ``Seek first to understand, then to be understood''
\item Clients and colleagues with whom one works are experts in their field or subfield who can teach statisticians something new. Instead of calling them ``clients,'' which might imply a relationship more transactional than collaborative, call them and treat them as ``domain experts.'' Domain experts are always worthy of respect.
\item Statisticians and data scientists play important roles in collaborative projects and are also always worthy of respect
\item A collaborative statistician is domain-expert centered, i.e., is committed to working with the domain expert who shows up---with all their strengths and faults---not just the idealized person one wishes had shown up
\item A collaborative statistician is ethical and abides by the ASA Ethical Guidelines for Statistical Practice \citep{COPE_ethical_2018}
\item It is possible and desirable for both domain expert and statistician to benefit from a collaboration. Statisticians should focus on creating shared goals with domain experts so that the outcome is a win for both parties.
\item Desirable outcomes are best achieved when statisticians commit to being helpful and to making a deep and long-term contribution by focusing on solving the domain problem---not just the statistics or data science questions.
\end{itemize}
The aspect of the problem on which the statistician focuses (i.e., the statistics problem or the domain problem) is a main differentiator between {\it consulting} and {\it collaboration}. Borrowing from definitions proposed by \citet{vance_recent_2015} and \citet{love_developing_2017}, we hold the following as working definitions:\\
\textbf{\emph{statistical consulting}} is ``working cooperatively with clients to answer statistics or data questions.''\\
\textbf{\emph{statistical collaboration}} is ``working cooperatively with domain experts to create solutions to research, business, and policy challenges and achieve research, business, and policy goals.''

Ultimately, as expressed by five winners of the American Statistical Association's W.J. Dixon Award for Statistical Consulting in \citet[p. 45]{love_developing_2017}, ``The principal determinant of the role of the statistician is the needs of the [domain expert]. An experienced statistician can continue to make valuable contributions at all levels, from providing basic answers to statistical questions to participating as a research team member to leading the way for other statisticians to do the same.'' We believe that the attitudes and behaviors described in the ASCCR Frame are useful for statisticians in all roles, including those of consultants, collaborators, and leaders.

\section{{\underline S}TRUCTURE}
\label{sec:structure}
Meetings, especially at the beginning of a project, are the primary mode for statisticians to interact with domain experts. According to \citet{snee_leading_2005}, meeting design and facilitation is an important skill for statisticians; the ability to run effective meetings is essential for teamwork because well-designed and well-run meetings help the team work together more effectively \citep{snee_non-statistical_1998}. Since statisticians often conduct many meetings---including one-on-one face-to-face meetings, phone and video meetings, and group meetings---we believe learning an effective structure for meetings will improve statisticians' ability to effectively collaborate. Project planning and management skills are also important aspects of the Structure of collaborations \citep{snee_non-statistical_1998}, and are beyond the scope of this article. 

Effective statistical collaboration requires focus on several things simultaneously, including: learning pertinent aspects of a new domain, considering the relevant statistical and data science issues of a problem, navigating the interpersonal dynamics and relationships between those present at the meeting, and organizing and managing the meeting. The use of checklists by surgeons has been shown to reduce the number of medical errors \citep{gawande_checklist_2009}. Airline pilots rely on well-designed checklists to improve safety \citep{degani_cockpit_1993}. We have found that learning and practicing a specific structure makes statistical collaboration meetings with domain experts and colleagues more effective and efficient. Transforming the organization and management of meetings into procedures that become routine habits frees statisticians' brains to focus on the ever-changing domain and statistical issues specific to the project and the relationship issues specific to the principals involved.

We have adapted the POWER Structure, created by Doug \citet{zahn_stumbling_2019}, to help us and our students structure statistical collaboration meetings. This structure reduces potential imbalances in who has the authority or control over a meeting by facilitating agreement by both statistician and domain expert on what will be mutually productive for the meeting and ensuring that important conversations occur. We practice this structure extensively through in-class exercises, role plays, observations, and Video Coaching and Feedback Sessions (VCFS) \citep{vance_using_2016} so that the organization and management of a meeting becomes routine, and we spend energy focusing on the other aspects of the collaboration.

\subsection{The POWER Structure}
POWER stands for Prepare, Open, Work, End, and Reflect, five components we believe should be present in every statistical collaboration meeting. These components are described below.
\subsubsection{Prepare}
Effective collaborators prepare for an upcoming meeting by becoming mentally, physically, and emotionally ready. Mental preparation includes reviewing the agenda, notes of previous meetings, and any materials the domain expert has sent. It also includes reviewing any unfamiliar terminology or statistical methods the domain expert referred to in their communications. Physical preparation includes arriving at the meeting room with sufficient time to arrange furniture as well as attending to physical needs prior to the meeting. Emotional preparation includes completing---before the meeting begins---unrelated tasks that might distract from the upcoming meeting, allowing one to focus on the domain expert and her research/business questions. We also recommend creating a flexible, domain-expert-centered plan for the meeting and writing key elements of such plan on a meeting room whiteboard or a page of notes for personal reference.

\subsubsection{Open}
We like to say, ``Win the opening, win the meeting. Win the meeting, win the collaboration,'' which is to say that, based on our experience, an effective opening sets the remainder of the meeting on a trajectory for success because it establishes the structure for the meeting. Similarly, an effective and efficient initial meeting enables the collaboration to flourish. A series of conversations at the beginning of a meeting are key to an effective opening, including:
\begin{itemize}
\item Greeting -- Introduce oneself and ensure that everyone in the room (or on the video/call) has been introduced; smile; make eye contact; shake hands; generally, help the domain expert feel comfortable, which could include engaging in small talk if mutually desired
\item Time conversation -- Check if the scheduled meeting time still works for everyone and whether they can stay longer if it would be useful. For example, ask: ``Does it still work for you to meet for [1 hour]? If we're being productive, for how long after [1 hour] could you stay? I could stay until [time]; I have another meeting in [1.5 hours].''
\item Short-term wanted conversation -- Ask what the domain expert would like to accomplish during this particular meeting. The statistical collaborator should consider adding her own goals to the domain expert's wants or even stating these goals before asking the domain expert. For example, consider saying, ``I would like to learn more about your research/business project because it helps me determine the best statistical analysis methods to use. What specifically would you like to accomplish during the [1 hour] we have for this meeting?''
\begin{itemize}
\item Paraphrase the domain expert's goals in one's own words, write them where everyone can see (e.g. a whiteboard), and ask, ``Is there anything else?''
\item In collaboration with the domain expert, prioritize the listed goals to create an initial agenda for the rest of the meeting
\item If an agenda for the meeting has already been created, review the agenda items to ensure they are still relevant and ask if anything should be added
\end{itemize}
\item Willing conversation -- Determine if one is willing to accomplish what the domain expert wants. Initially determine this for oneself and then communicate this to the domain expert. This can occur later in the meeting when one has a better understanding of what is needed to accomplish the meeting goals.
\item Able conversation -- Determine if one is able to accomplish all that the domain expert wants. This also can occur later in the meeting. The statistician should communicate her willingness and ability to proceed in the collaboration.
\item Medium- and Long-term goals and timelines -- Ask what the domain expert would like to accomplish during the medium-term (e.g., by the end of the semester or the fiscal quarter) and the long-term (e.g., 1+ years) of her project. Ask about any upcoming deadlines and the overall timeframe of the project.
\end{itemize}
A common excuse for not conducting these opening conversations is that, ``They take too much time.'' In our experience, these conversations are crucial for effective collaborations and having such conversations early in the collaboration actually saves time.

\subsubsection{Work}
After these opening conversations are completed, we recommend transitioning into the working phase of the meeting by asking the domain expert to explain in more detail her overall research or business goals. Essentially, the entire ASCCR Frame applies to the working phase of every meeting. With a collaborative attitude, effective statisticians operate within the structure they have created for the meeting to learn about the domain expert's project and work with her to address her short-term content goals for the meeting and medium- and long-term content goals for the project while communicating effectively and building or strengthening the relationship.

\subsubsection{End}
From our extensive experience, statistical collaboration meetings too often end in a rush, and decisions made or next steps to be taken are forgotten unless sufficient time is dedicated to summarizing the meeting, outlining the next steps for the project, and recording the summary and next steps in writing. We recommend reserving and dedicating 10--20\% of the meeting time to have a complete ending of the meeting (e.g., 10 minutes for a 60-minute meeting).

During the ending, summarize all major decisions made. Ask the domain expert, ``What can I clarify?'' Specifically review the domain expert's initial goals for the meeting and what the outcome was. Ask if the domain expert's goals were satisfactorily accomplished. If a goal was not satisfactorily addressed, devise a plan to address it (e.g., schedule another meeting or commit to addressing the issue over email).

The statistician should clarify with the domain expert any deadlines for the project and discuss a plan. Both parties should agree on a timeline for this plan. Specifically, the statistician and domain expert should agree on who will carry out each item of the plan, when it will be done, and how they will communicate to the other that it was done. The end of a meeting is a great time to schedule a follow-up meeting or decide how and when a follow-up meeting will be initiated.

Finally, the statistician should compose a meeting summary report that records the decisions made, the next step action items, the timelines for these actions, and how or when the next meeting will be scheduled. The statistician should ideally send this meeting summary to the domain expert immediately after the meeting and solicit corrections or additions to the summary. Sometimes the statistician will write all of the salient points on a whiteboard during the meeting, in which case sending a photo of the whiteboard to the domain expert might suffice as a written meeting summary report.

\subsubsection{Reflect}
To gain the most benefit from the POWER Structure, we believe it is imperative for the statistician to reflect on what aspects of the meeting and the structure went well and where there may be opportunities for improvement. After the meeting has been summarized and ended, it may be beneficial to ask the domain expert to help in this reflection. One might ask, ``What aspects of today's meeting did you find especially effective? What opportunities can you see to improve the way our meetings are structured in the future?''

According to \citet{mann_reflection_2007}, reflective capacity is regarded by many as an essential characteristic for professional competence. \citet{epstein_defining_2002} include the use of reflection in daily practice in their definition of professional competence for physicians. \citet{schon_reflective_1983} concluded that through a feedback loop of experience, learning, and practice, professionals can continually improve their work. Engaging in reflection has been shown to improve learning and skill development for nurses \citep{glaze_reflection_2001}, organizational leaders \citep{nesbit_role_2012}, and statistics students \citep{hall_improving_2010}. Statisticians and data scientists can also employ reflection to improve their collaboration skills.

\subsection{The POWER Structure in Practice}
We recommend recognizing which elements of the POWER Structure one already implements in statistical collaboration meetings; then identifying a single unimplemented element of the structure to incorporate into one's next meeting and reflecting on how it went. One can continue adding elements of POWER to meetings and reflecting on the effectiveness of these actions or behaviors. For example, a first step toward implementing the POWER Structure would be to spend five minutes preparing oneself mentally, physically, and emotionally for an upcoming meeting and then reflecting on if the meeting went any better or worse as a result. A follow-up step for a subsequent meeting would be to again prepare for it, and then open the meeting by verifying the time available and explicitly asking the domain expert what she would like to accomplish during that meeting. Referring back to these initial goals throughout the meeting and explicitly summarizing what was decided for each of these goals would complete this next step toward implementing the POWER Structure.

We have found that practicing the conversations embedded in the POWER Structure---especially those in Open---with a friend or colleague outside of a meeting setting (e.g., during a role play exercise) can be helpful. Hearing how the words sound out of one's own mouth or from a colleague can make performing these tasks in a real meeting much easier.

\section{{\underline C}ONTENT}
We believe that the Content of every statistics or data science project has three components: Qualitative, Quantitative, and Qualitative. These components generally follow the order of beginning, middle, and end. Every effective collaboration must start with the Qualitative ($Q1$) aspects of the project and must also end with the Qualitative ($Q_3$). A mistake many statisticians and data scientists commit, especially beginning ones, is jumping into the Quantitative ($Q_2$) aspects of a project too early, i.e., before they have verified their understanding with the domain expert of the Qualitative ($Q_1$) aspects of the project or problem \citep{kimball_errors_1957}. Another framework for thinking about the Content of a collaboration in the lifecycle of a data science project is the PCS (predictability, computability, and stability) framework of \citet{yu_three_2019}. Here we describe the Qualitative-Quantitative-Qualitative ($Q_1Q_2Q_3$) process we have adopted and adapted from \citet{leman_developing_2015} to help guide statisticians and data scientists in navigating the content of every collaboration project.

\subsection{Qualitative ($Q_1$)}
$Q_1$ is the project's initial Qualitative component, which sets the foundation for the Quantitative component of the project ($Q_2$) and the implementation of the solution ($Q_3$). Specifically, there are seven aspects of $Q_1$, inspired by the questions of George Heilmeier's Catechism \citep{hahn_engineering_2014}, relevant for every applied statistics project. The information for each aspect should be conveyed and that information should be verified between the statistician and the domain expert.

1. {\it What is the domain problem?} It may be that a collaboration will address several domain problems or that the initial domain problem identified turns out not to be one of the problems addressed. It is often the case that the process of collaboration helps the domain expert refine the statement of her domain problem or even identify a wholly new domain problem to be solved. Examples of a statistician becoming a lead author on a paper by identifying a new domain problem to solve include \citet{vance_computed_2013} and \citet{stallings_determining_2013}.

2. {\it Why is the domain problem important or interesting?} Understanding the reasons the domain expert finds the problem important and/or interesting can clarify the problem and help the statistician identify the best quantitative ($Q_2$) methods for solving it. Reaching this level of understanding about the motivations of the domain expert to solve the domain problem can be accomplished by asking a series of ``Why?'' questions, preceded by a statement of the statistician's intent in asking these seemingly impertinent questions.

3. {\it How will the eventual solution be used?} Eliciting this information can clarify a complicated project and identify the implementation of the solution as an ultimate goal toward which to work. This information can also inform future decisions about which statistical methods ($Q_2$) to use.

4. {\it What potential data could solve the domain problem? I.e., what data, if it were available and accessible, would help answer the underlying research questions or guide the business or policy decisions?} This is an important theoretical exercise that is too often skipped in statistics and data science collaborations. It may be that the solution to the domain problem becomes simplified by collecting new data and that the domain expert and data scientist overlook this by focusing only on data already collected. On the other hand, the impossibility of collecting certain data can illuminate the intricacies of the domain problem and help guide the collection or analysis of alternative data.

5. {\it What data have been collected? Why were the data collected originally? For what purpose? When and where were the data collected? Who or what collected the data? How were the data collected, i.e., with what instrumentation/methods?} Understanding the data and their collection will help the statistician appropriately explore, visualize, model, and analyze the data \citep{chatfield_problem_1995}.

6. {\it What may be the qualitative relationships between variables, for those observed and unobserved?} Many of the domain problems statisticians address are causal in nature, and domain experts often want to be able to make causal claims when implementing the solution. Sketching a casual diagram \citep{pearl_causal_1995} alongside the domain expert can be very useful when discussing potential models.

7. {\it Which types of statistical analyses or machine learning techniques would be most useful to the domain expert? Which would not be useful? Could new statistics or data science methods be developed to more usefully answer the domain question?} If the domain expert does not understand or cannot defend the choice of a specific technique, that technique may not be useful, and a different (simpler or more familiar) method may need to be used \citep{stallings_type_2014}. If the simpler method is not appropriate for the problem and the data, the more advanced method should be used and sufficient time spent explaining the new method to the domain expert. Statisticians and data scientists have the potential to drive collaborative projects to develop new technical methods tailored to better answer the particular domain questions and generalizable to answer other domain questions. Examples include \citet{vance_social_2009} and \citet{mullen_mapping_2019}.

\subsection{Quantitative ($Q_2$)}
Technical skills in statistics and data science and how to effectively apply them are the \emph{sine qua non} of statistical collaboration. Applying these skills to quantitatively explore, visualize, model, and analyze data to solve a research/business/policy problem in a reproducible manner is the Content of collaborations that occurs after the problem has been well defined and understood ($Q_1$) and before the solution to that problem is determined, explained, and implemented ($Q_3$). Keeping a reproducible record of decisions made on how the raw data have been tidied and analyzed is essential when there is the potential for more than one person to work on the data.

These quantitative skills are the technical core of the $Q_1Q_2Q_3$ process and, therefore, of statistical collaboration. Textbooks on statistical consulting often focus on this quantitative ($Q_2$) content \citep{cabrera_statistical_2002, hand_statistical_2007} as do many university courses in statistical consulting \citep{jeske_merging_2007, khachatryan_incorporating_2015}. In our experience, effective statistical collaboration must begin with strong foundational knowledge of the theory and methods of statistics and data science. Strong technical skills coupled with effective professional skills produce strong collaboration skills.

\subsection{Qualitative ($Q_3$)}
Translating the ($Q_2$) data analysis into useful qualitative answers to the original research question or the implementation of a data-driven decision completes the $Q_1Q_2Q_3$ process. A statistician who sufficiently understands the domain problem ($Q_1$) and appropriately analyzes the data ($Q_2$) is not finished until she explains to the domain expert---qualitatively---what the analysis results mean ($Q_3$). A collaboration is sufficiently completed ($Q_3$) only after the statistician and domain expert jointly interpret the quantitative results ($Q_2$) to answer the original research question ($Q_1$) (see Section 5.3 for a framework for explaining statistics to nonstatisticians). Additional aspects of $Q_3$ include visually displaying and communicating the results of the analysis, detailing the conditions or assumptions made in the quantitative analyses and the resulting limitations of the conclusions from the analyses, and working with the domain expert to effectively explain these results and the conclusions to her advisor/boss/peers, perhaps through coaching and/or feedback during a meeting. \citet{murdoch_interpretable_2019} discuss a framework for interpreting outputs of machine learning algorithms. Ultimately, the final part of the Content of a collaboration is using statistics and data science to communicate the narrative of the data and useful answers to the research, business, or policy questions.

\section{{\underline C}OMMUNICATION}
Communication is, of course, key to effective interdisciplinary collaborations as described in past literature on statistical consulting and collaboration \citep{boen_human_1982, hoadley_communications_1990, derr_statistical_2000}. We have developed material specific for statisticians and data scientists to learn and practice three important aspects of communication, which we put into a framework called the Triangle of Statistical Communication. These aspects are: Asking Great Questions; Listening, Paraphrasing, and Summarizing; and Explaining Statistics to Nonstatisticians. We provide an overview of these aspects of communication below. Other important aspects of communication in statistical collaborations not described here include nonverbal communication, communicating statistical concepts and results orally and in writing (e.g., giving presentations and writing interim and final reports), data visualization, and providing useful feedback.

\subsection{Asking Great Questions}
According to our model for asking great questions \citep{vance_collaboration_2018}, a question is ``great'' is if it accomplishes two aims: 1) The question \emph{creates shared understanding} between the statistician and domain expert by eliciting information useful for answering the research/business/policy questions, and 2) The question is asked in a way that \emph{strengthens the relationship} with the domain expert. Great questions can be asked in all five aspects of the ASCCR Frame. For example, a great question regarding Communication within a collaboration is, ``I'm not quite sure I completely understood everything you said. Would it work for you if I restate it and you can tell me what I'm missing?''

\subsection{Listening, Paraphrasing, and Summarizing}
When a domain expert answers questions, the statistician must, of course, listen to the domain expert to \emph{understand} the content of the response, paraphrase the response to \emph{clarify} content, and then summarize to \emph{solidify} the content. As with asking great questions, effective listening, paraphrasing, and summarizing achieves two aims: creating shared understanding and strengthening the relationship between statistician and domain expert. Here are six tips and strategies for listening, paraphrasing, and summarizing:
\begin{itemize}
\item Focus as much as possible on the domain expert while listening. Manage distractions. Be patient.
\item Listen to the domain expert's words without jumping ahead to their implications
\item Paraphrase whenever the domain expert communicates a key concept or whenever one is confused or uncertain
\item State the intent behind attempts to paraphrase and summarize (e.g., ``Let's see if I understood correctly; you are attempting to\ldots'')
\item Use visual aids (e.g., a whiteboard) when paraphrasing and summarizing so the domain expert can both hear and see the restatement or summary
\item Summarize conversations completely before moving on to the next topic.
\end{itemize}

\subsection{Explaining Statistics to Nonstatisticians}
We have adopted and adapted Azad's ADEPT method \citeyearpar{azad_math_2015} to explain statistical terms and concepts to nonstatisticians using Analogies, Diagrams, Examples, Plain language, and Technical definitions (ADEPT). Using analogies as a bridge between familiar situations and new situations (i.e., relating a concept already known to the nonstatistician to help explain an unknown statistical concept) can be extremely effective \citep{martin_its_2003}. Such analogies can help scaffold learning \citep{holyoak_mental_1995, podolefsky_analogical_2007} by providing context for a new idea and the general form of the concept, with details supplied as needed. Effective analogies are, however, quite difficult to generate spontaneously. Therefore, we recommend developing and practicing suitable analogies for concepts that arise most often in statistical collaborations before they are needed and then deploying them as appropriate to explain concepts such as power and sample size calculation, training and test datasets, types I and II errors, p-values, confidence intervals, residuals, correlation, sampling, inference, and multiple comparisons.

Diagrams are also extremely effective tools for explaining statistical concepts, especially for visual learners. Like analogies, diagrams must also be carefully designed so as to clarify rather than blur the concept. Examples are most effective for explanations when drawn from the nonstatistician's experience using words and ideas from her domain or everyday life. Similarly, the best explanations use plain language, avoiding jargon that must also be explained.

To effectively explain a statistical concept, not all five aspects of ADEPT must be used. For example, providing the technical definition might illuminate a concept for one nonstatistician but obscure it for another. Mathematically inclined domain experts often appreciate seeing the technical definition of a new statistical concept. However, in our experience, providing a technical definition to complete an explanation is unnecessary and unhelpful for most domain experts. Technical definitions should therefore be used with discretion, perhaps preceded by an inquiry about its possible usefulness (e.g., ``I could write down the mathematical definition of this concept. Would that be useful to you?'')

\section{{\underline R}ELATIONSHIP}
Developing strong relationships with domain experts is key for statisticians and data scientists to become effective collaborators because, according to Zahn, ``Relationships are critical to success no matter what you do'' \citep[p.41]{love_developing_2017} and according to Boroto and Zahn, ``The bulk of the statistician's job is communication for the purpose of building effective relationships'' \citeyearpar[p. 72]{boroto_becoming_1989}. To develop strong relationships, we apply the concepts of social awareness and social management from the emotional intelligence literature \citep{goleman_social_2006}. We believe that collaborative statisticians should attempt to become more aware of the domain experts' attitudes and emotions and should choose their own words and actions to manage/strengthen the relationship.

Some practical advice on how to build stronger relationships with domain experts includes:
\begin{itemize}
\item Express authentic interest in developing and building a strong relationship with the domain expert
\item Learn and use the language of the domain expert's discipline
\item As mentioned in the section on Attitude, respect the skills the domain expert brings to the collaboration because strong relationships are built on respect
\item Be aware that some relationships might not be worth continuing. If a domain expert does not exhibit respect for the statistician, it may be worthwhile to end that relationship and focus on strengthening others.
\item Building strong relationships requires time, patience, and trust. Act trustworthy to gain trust.
\end{itemize}

Situational leadership is a theory of leadership recognizing task and relationships as important dimensions of leader behavior \citep{hersey_life_1969, hersey_management_1977}. In some situations, the most effective leader has both high task-orientation and high relationship-orientation toward her followers \citep{hersey_revisiting_1996}. A unifying theme for our theory of collaboration as encompassed by the ASCCR Frame is that the actions of collaborative statisticians can be both task- and relationship-oriented. We believe that the most effective collaborations occur when the statistician focuses on: 1) helping achieve the shared goals of the collaboration (task behavior) and 2) strengthening the relationship with the domain expert (relationship behavior). When a statistician chooses an action that does both, for example when asking a great question, information is elicited to help accomplish the task AND the relationship is strengthened.

\section{IMPLEMENTING THE ASCCR FRAME INDIVIDUALLY AND IN THE CLASSROOM}
We believe that the ASCCR Frame can be useful for individuals at all career levels, for teachers of statistics and data science, and for organizations interested in improving the collaboration skills of their statisticians and data scientists. The first steps to implement the ASCCR Frame will differ based on the individual's experience. Beginning statisticians (students and early-career professionals) should focus on different aspects of ASCCR than experienced statisticians (mid- and late-career professionals). Below we outline five steps for beginning statisticians and three steps for experienced statisticians to guide their initial implementation of the ASCCR Frame to improve their collaboration skills and thus have greater impact in their organizations. We also briefly discuss how to teach the ASCCR Frame in the classroom.

\subsection{How Beginning Statisticians Can Implement ASCCR}
{\it Step 1: Adopt an Attitude of Collaboration}\\
We recommend beginning statisticians start with a clean slate and adopt the attitude that:
\begin{itemize}
\item The statistician enters a collaboration with curiosity to learn about the domain problem
\item The statistician will apply what she knows about statistics and data science to help solve the problem and implement a solution
\item The statistician will \emph{expand} her statistics and data science knowledge to solve the problem and implement a solution
\item The statistician's overall goals are to help the domain expert succeed and expand her knowledge and capabilities in the process. In other words, with an attitude that both can learn from this experience, the statistician will aim for a collaborative win-win outcome.
\end{itemize}

{\it Step 2: Implement Structure for Collaboration Meetings}\\
Entering each collaboration meeting with a plan and a proposed structure will help increase the effectiveness of each meeting. As described in Section 3.2, we recommend implementing one aspect of the POWER structure at a time and reflecting on how it went.

{\it Step 3: Commit to Completing $Q_1$ Conversations Before Moving to $Q_2$}\\
To reduce the chances of rushing the $Q_1$ conversations and jumping into $Q_2$ too quickly, we recommend stating the intent behind asking the domain expert the $Q_1$ questions in Section 4.1 and then summarizing the responses before transitioning into $Q_2$. For example, ``Before we jump into the quantitative content of your project, it will really help me to understand the qualitative aspects of your project so we can decide what the best statistical methods may be to answer your research/business/policy questions. Will that work for you?''

{\it Step 4: Improve Communication Skills}\\
Before a collaboration meeting, transform a few commonly asked questions into great questions and write them down on a sheet of notes. Ask these great questions in the meeting; listen, paraphrase, and summarize the responses; and reflect on how well these questions and your listening, paraphrasing, and summarizing elicited useful information for the project and helped strengthen the relationship with the domain expert. Continue improving communication skills by identifying one statistical concept frequently explained and transforming this explanation, before a meeting, using the ADEPT method.

{\it Step 5: Cultivate Strong Relationships}\\
Express interest in cultivating a strong relationship with a new or existing domain expert. For example, begin a meeting by saying, ``I'm really looking forward to working with you on this project because [insert authentic reasons].'' Try ending a meeting by asking the domain expert, ``As you know, I'm new to [this project, this company, statistical collaboration]. I would really like to improve my collaboration skills so I can help you achieve your [research/business/policy] goals. What could I do differently next time to improve our meetings or the way we work together?''

\subsection{How Experienced Statisticians Can Implement ASCCR}
{\it Step 1: Reassess Attitude}\\
We recommend experienced statisticians recommit themselves to the following goals:
\begin{itemize}
\item ``Seek first to understand, then to be understood'' (aka \emph{The Fundamental Law of Statistical Collaboration})
\item Being helpful to the domain expert is the main goal
\item Identify how one can make a deep contribution to the research/business/policy and seek to make such a contribution
\item Smile more :) Express your enthusiasm for the project to make engaging with statistics and data science a fun experience for all.
\end{itemize}

{\it Step 2: Improve Communication Skills}\\
The questions we ask and the way we explain statistics can always be improved. We recommend intentionally transforming good questions into great questions and adding more aspects of the ADEPT method into future explanations.

{\it Step 3: Cultivate Stronger Relationships}\\
Asking domain experts the following question at the end of collaboration meetings can sometimes reveal actionable information and will almost certainly cultivate stronger relationships. Ask, ``I've been doing this for a long time, but that doesn't mean I can't still learn and improve my collaboration skills. What could I do differently next time to improve our meetings or the way we work together?''

\subsection{Teaching ASCCR in the Classroom} 
We teach interdisciplinary collaboration and statistical consulting in dedicated semester and quarter-long courses at the graduate and undergraduate level, and the ASCCR Frame provides a unifying structure for these courses. We teach the components of ASCCR separately as well as explore with the students how a particular cross-cutting topic might relate to two or more components of ASCCR. An example of the former is that in a class period dedicated to Structure, we review the POWER structure, model for the class what an effective Opening of a meeting looks like through a live role play and/or video, coach students in their own repeated role plays as they practice various aspects of Opening a meeting, and engage students in reflecting on this experience.

An example of a cross-cutting topic addressing at least four components of ASCCR is the statistical analysis plan (SAP), in which the statistician sends to the domain expert a written summary of her understanding of the problem ($Q_1$); a plan for tidying, exploring, modeling, and analyzing the data ($Q_2$); and some sample interpretations of hypothetical results ($Q_3$). The SAP can provide helpful Structure for the project, follows the $Q_1Q_2Q_3$ process (Content), and is a key element of Communication as the SAP can be transformed into a final report with the addition of results of the planned analyses. An instructor might also emphasize how an effective SAP might strengthen the Relationship with the domain expert by expressing authentic interest in the project, demonstrating proficiency using the language of the domain expert's discipline, and establishing trust by honoring commitments.

Since collaboration is so fundamental to the work of a statistician or data scientist, components of ASCCR should be integrated into other courses, such as an undergraduate applied modeling course. Explicitly introducing the $Q_1Q_2Q_3$ process (Content) as a guiding framework for any applied problem can help students think beyond the technical modeling procedures to what question they are answering and what the model implies is the answer to that question. Students can also be taught ethics of analyzing data and communicating results \citep{lesser_ethical_2004}, which is an important Attitude of collaboration. After each assignment or project, students might be asked to reflect on an ethical component of the project, such as including a paragraph in their report discussing who might benefit and who might suffer as a result of the ($Q_3$) implementation of a particular analysis. If group work is a component of the course, students can be taught how to provide helpful feedback (part of Communication), how to create reproducible workflows (part of $Q_2$ of Content), and how to build strong Relationships with classmates. Finally, when a new statistical concept is introduced, the instructor may use the ADEPT method (Communication) to help students understand the concept. Students might also be asked to explain course concepts to a classmate using aspects of ADEPT, thereby increasing their own understanding.

\section{CONCLUSION}
\label{sec:conc}
Collaboration skills are learnable professional skills that can improve the effectiveness of statisticians and data scientists and thereby increase their impact. We have provided an overview of the five components of Attitude, Structure, Content, Communication, and Relationship that comprise the ASCCR Frame for Collaboration and specific steps beginning and experienced statisticians can take to begin improving their collaboration skills. While this framework is general and could be applied by those who collaborate in any field, we focused specifically on applying this framework to collaborative statisticians and data scientists. We intend for the ASCCR Frame to help organize and stimulate research and teaching in interdisciplinary collaboration. As such, we call on the statistics and data science communities to explore how new or existing research in and teaching methods for collaboration can be incorporated into the ASCCR Frame and how evidence can be generated to test the effectiveness of the ASCCR Frame for improving statisticians' and data scientists' collaboration skills. We plan to provide more detail on the five components of ASCCR in a series of future publications and create more materials for learning and teaching collaboration to augment those already posted on our \emph{Collaboration in a Bag} public repository at \url {www.osf.io/xmtce} for individuals to learn effective collaboration skills and organizations to teach them \citep{vance_public_2019}.

\bibliographystyle{natbib}  
\bibliography{ASCCR}
\end{document}